# MATT-GS: Masked Attention-based 3DGS for Robot Perception and Object Detection

Jee Won Lee[1,2], Hansol Lim[1,2], SooYeun Yang[1,2], and Jongseong Brad Choi[1,2*]

*Abstract*—This paper presents a novel masked attention-based 3D Gaussian Splatting (3DGS) approach to enhance robotic perception and object detection in industrial and smart factory environments. $U^2$-Net is employed for background removal to isolate target objects from raw images, thereby minimizing clutter and ensuring that the model processes only relevant data. Additionally, a Sobel filter-based attention mechanism is integrated into the 3DGS framework to enhance fine details—capturing critical features such as screws, wires, and intricate textures essential for high-precision tasks. We validate our approach using quantitative metrics, including L1 loss, SSIM, and PSNR, comparing the performance of the background-removed and attention-incorporated 3DGS model against the ground truth images and the original 3DGS training baseline. The results demonstrate significant improvements in visual fidelity and detail preservation, highlighting the effectiveness of our method in enhancing robotic vision for object recognition and manipulation in complex industrial settings.

## I. INTRODUCTION

In recent years, the emergence of smart factories and advanced robotics has revolutionized industrial automation, significantly enhancing efficiency, precision, and safety in manufacturing environments [1], [2]. As production systems evolve, the integration of intelligent robots capable of complex perception and manipulation tasks has become crucial. In smart factories, where variability and high-speed operations are ordinary, robots must accurately identify and interact with numerous components even those as minute as screws and wires to ensure quality and reliability in assembly and maintenance processes [3]. Traditional approaches to robotic perception often rely on basic 2D imaging or coarse 3D representations, which struggle with cluttered environments and lack the resolution needed to capture fine details [4]. These limitations can result in misidentification or errors in manipulation, leading to inefficiencies and increased operational costs [5]. Furthermore, while simulation environments and digital twins have become indispensable tools for designing and testing robotic systems, they too suffer from insufficient detail and robustness when representing complex scenes [6].

Modern technologies such as 3D modeling and image-based reconstruction offer a promising solution for addressing these challenges. Methods such as Structure-from-Motion (SfM) [7], Multi-View Stereo (MVS) [8], and Neural Radiance Fields (NeRF) [9] have been effective in generating detailed 3D representations of complex environments. These techniques enable inspectors and robotic systems to visualize, analyze, and document environments with clarity, providing a viable alternative to traditional methods. However, while effective, these approaches have notable limitations. SfM and MVS often struggle with irregular or large-scale geometries and require extensive computational resources, while NeRF, despite its ability to generate highly detailed models, is typically unsuitable for real-time applications.

To overcome these challenges, 3D Gaussian Splatting (3DGS) has emerged as a superior alternative, offering both computational efficiency and the ability to handle unstructured geometries without the need for detailed surface meshes [10]. Compared to NeRF, 3DGS is computationally faster and demands significantly fewer resources, making it particularly well-suited for large-scale applications in dynamic environments. Nevertheless, standard 3DGS methods can sometimes miss fine details that are critical for high-precision tasks, such as the subtle features of screws, wires, and other small components.

To address these shortcomings, we propose a novel masked attention-based 3D Gaussian Splatting (3DGS) model that leverages advanced background removal and fine-detail enhancement techniques. By employing $U^2$-Net for precise segmentation, our approach isolates objects from background noise, ensuring that the reconstruction process focuses exclusively on relevant data [11]. Additionally, by reducing the number of key points during the structure-from-motion process, we noticeably reduce both the training time and memory usage associated with Gaussian splatting. Moreover, the integration of a Sobel filter-based attention mechanism within the 3DGS framework enhances the representation of critical details, thereby capturing intricate features that are essential for reliable object detection and robotic manipulation. Our method offers significant improvements over conventional techniques by delivering high-fidelity 3D reconstructions that increase detection accuracy and facilitate more reliable simulation and manipulation in smart factory settings. The enhanced visual fidelity achieved through our approach underscores its potential to advance robotic perception in both real-world and simulated environments. Ultimately, this work paves the way for more efficient, accurate, and robust systems capable of meeting the rigorous demands of modern automation.

Jee Won Lee[1,2], Hansol Lim[1,2], SooYeun Yang[1,2], and Jongseong Brad Choi[1,2*] are with [1]Mechanical Engineering, State University of New York, Korea, 21985, Republic of Korea, [2]Mechanical Engineering, State University of New York, Stony Brook, Stony Brook, NY 11794, USA (e-mail: jeewon.lee@stonybrook.edu; sooyeun.yang@stonybrook.edu; hansol.lim@stonybrook.edu; jongseong.choi@stonybrook.edu).

This work is supported by the National Research Foundation of Korea 14 CHOI et al. (NRF) grant funded by the Korea government (MSIT) (No. 2022M1A3C2085237).
(*corresponding authors: Jongseong Brad Choi).

## II. RELATED WORKS

### A. U-Net

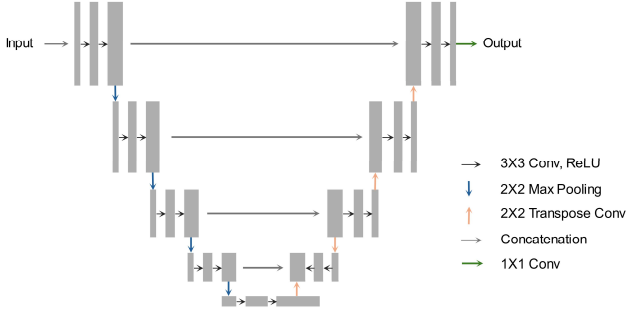

Fig. 1. Free Body Diagram of EV during motion

U-Net, introduced by Ronneberger et al. in 2015, has become a cornerstone in image segmentation, particularly within biomedical applications. Its architecture follows an encoder-decoder structure. The encoder captures context through a series of convolutional and pooling operations, while the decoder enables precise localization by performing upsampling and concatenating corresponding high-resolution features through skip connections. Mathematically, the convolution operation at layer $l$ can be expressed as:

$$f_{l+1} = \sigma(W_l * f_l + b_l) \tag{1}$$

Where $f_1$ is the feature map at $l$, $W_l$ is the convolution kernel, $b_l$ is the bias, and $\sigma$ represents a nonlinear activation function, typically ReLU.

In order to recover the spatial details that are lost during downsampling, skip connection are essential as they combine the high-resolution features from the encoder with the upsampled features from the decoder. The final segmentation output is generated by applying a $1 \times 1$ convolution followed by a softmax activation. For training, a combination of the cross-entropy loss and the Dice coefficient loss is often used, where the Dice loss is given by:

$$L_{Dice} = 1 - \frac{2\Sigma_i p_i g_i}{\Sigma_i p_i + \Sigma_i g_i} \tag{2}$$

Where $p_i$ and $g_i$ represents the predicted and ground truth labels, respectively.

$U^2$-Net, proposed by Qin et al. in 2020, builds on the foundational U-Net architecture to address its limitations in capturing fine-grained details in complex scenes [12]. $U^2$-Net introduces a nested U-structure within its network, known as Residual U-blocks (RSU). Each RSU block functions as a mini U-Net, performing hierarchical feature extraction at multiple scales. This nested design allows the network to aggregate multi-scale contextual information more effectively, which is crucial for segmenting subtle structures. A simplified formulation of an RSU block's output is as following:

$$Y_{RSU} = F(X) + X \tag{3}$$

Where $X$ is the input feature map, and $F(X)$ represents the transformation within the nested U-structure.

The residual connection, $X$, aids in gradient propagation during training. Moreover, $U^2$-Net generates side outputs at various stages of the network, which are fused to produce the final segmentation map:

$$\hat{Y} = \sum_{i=1}^{N} \alpha_i Y_i \tag{4}$$

$Y_i$ are the side outputs from different scales, and $\alpha_i$ are learnable weights that determine the contribution of each side output.

The progression from U-Net to $U^2$-Net represents a significant enhancement in the ability to capture fine details and multi-scale information. While U-Net established a robust framework for segmentation with its encoder-decoder structure, its relatively shallow design can be limited when precise segmentation of small or complex objects is required. $U^2$-Net overcomes these limitations through its nested architecture, which better preserves detailed structures in the segmentation process.

### B. Attention Mechanism

In image processing context, "attention" refers to techniques that enhance and emphasize important features such as edges, textures, or fine details in an image. Although the term is also widely used in deep learning, the attention mechanism discussed here is distinct in that it focuses on amplifying spatial detail and contrast within a given image or feature map.

Early work in image processing recognized the need to highlight salient image features. Classical approaches, such as visual saliency models, aim to mimic human visual perception by identifying regions with high contrast or distinctive structural information [13]. These models often compute local gradient magnitudes or other contrast-based measures to generate an "attention map" that indicates regions likely to contain important details. Given a feature map F, an attention map $A$ can be defined as:

$$A = \phi(F) \tag{5}$$

Where $\phi$ is an operator that extracts features such as gradient magnitude or local contrast. The enhanced feature map $F'$ is then computed as:

$$F' = F + \lambda \cdot (A \circ F) \tag{6}$$

Where $\lambda$ represents a scaling factor and $\circ$ denotes element-wise multiplication. This formulation allows the model to improve the representation of regions with strong gradients or high contrast, ensuring that fine details are more pronounced.

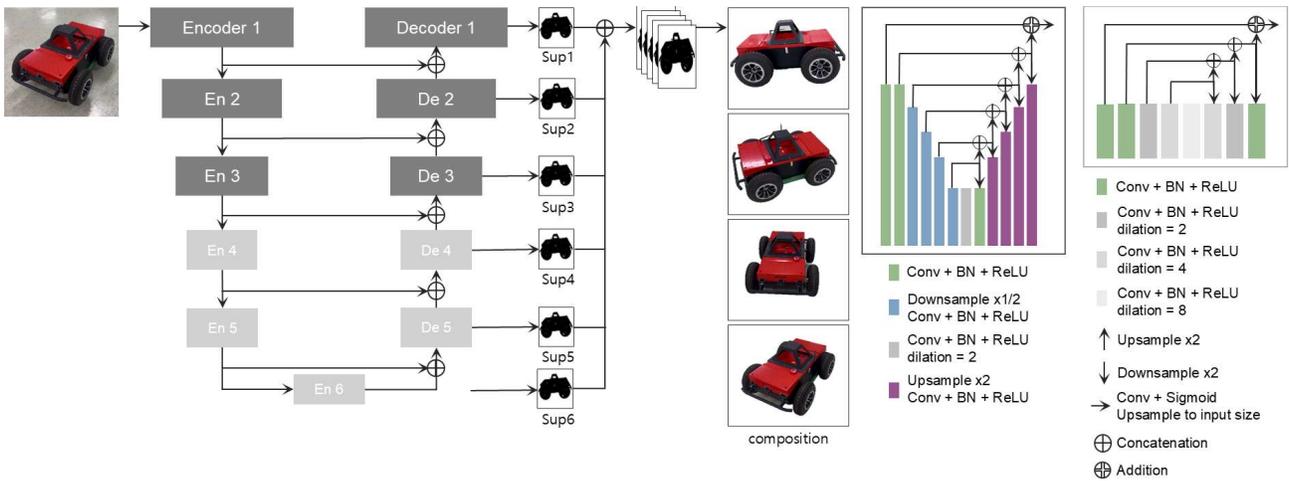

Fig. 2. An Overview of the $U^2$-Net architecture. This figure illustrates the encoder-decoder design and nested U-blocks and multi scales of feature extraction and skip connections combine to preserve fine details while capturing global context.

### C. 3D Gaussian Splatting

3D Gaussian Splatting (3DGS) builds upon the sparse point cloud generated through Structure from Motion (SfM) by producing a more detailed and visually realistic representation. Instead of using conventional mesh-based models, 3DGS utilizes Gaussian splats, ellipsoidal representations of color and opacity, to reconstruct a scene in a photorealistic manner. For each Gaussian splat, four key attributes are defined: position, covariance matrix, opacity, and color. The position of each splat is represented by a vector x containing its x, y and z coordinates, while the covariance matrix $\Sigma$ governs its rotation and scaling, defining the spatial extent of the ellipsoid. This covariance matrix is mathematically formulated as:

$$\Sigma = RSS^TR^T \qquad (7)$$

where $S$ represents the scaling matrix controlling the size of the ellipsoid along different axes, and $R$ denotes the rotation matrix, aligning the splat within the 3D coordinate space. These parameters collectively determine the shape and orientation of each splat, allowing them to be effectively combined to form a complete 3D reconstruction. The probability density function of the Gaussian splat is given by:

$$G(x) = e^{-\frac{1}{2}(x)^T\Sigma^{-1}(x)} \qquad (8)$$

where the opacity value $\alpha$ is scalar ranging from 0 to 1, controlling the transparency of each splat. The color of a Gaussian splat is represented using Spherical Harmonics (SH), which model the directional color distribution of the radiance field based on viewing angles.

To enhance the quality of the reconstructed scene, the gaussian splats undergo iterative optimization by comparing rendered views to actual camera perspectives. By continuously refining the splat attributes through multi-view consistency checks, the 3DGS model can generate highly realistic visualizations at an efficient computational cost. Compared to traditional mesh-based approaches, 3DGS offers superior flexibility in representing complex and unstructured environments.

### III. METHODOLOGY

#### A. Overview

To generate a masked high-fidelity 3D model, we first segment and remove the background clutter using $U^2$-Net and reduce the key points during the structure-from-motion process to streamline training. Then an attention-enhanced 3D Gaussian Splatting module reconstructs detailed 3D representations with preserved fine features.

#### B. Background Removal and Segmentation

Robust background removal is essential for isolating objects from irrelevant scene elements and ensuring that subsequent processing focuses on the regions of interest. Inspired by the $U^2$-Net architecture [14], which utilizes a nested U-structure to capture multi-scale contextual information while preserving fine details, a similar approach is employed for segmentation in this work.

A custom training dataset was assembled, consisting of images captured from various robotic devices and sensors to reflect diverse operational conditions. Data augmentation techniques including random rotations, horizontal and vertical flips, and scaling were applied during training to enhance the model's robustness to variations. Additionally, the input resolution was increased by rescaling images to 512 pixels using the RescaleT transformation, which helps preserve finer details, which helps preserve finer details, although at the cost of increased computational demand.

In practice, the segmentation process first generates a salient map that highlights the object of interest. This saliency map is normalized to a [0, 1] range and then converted into a binary mask through thresholding. By compositing the original image with a white background using the binary mask, the final output retains the segmented object while rendering the rest of the scene in white. This procedure not only enhances the clarity of the object boundaries but also provides high-quality, detail-rich inputs for subsequent stages in the processing pipeline.

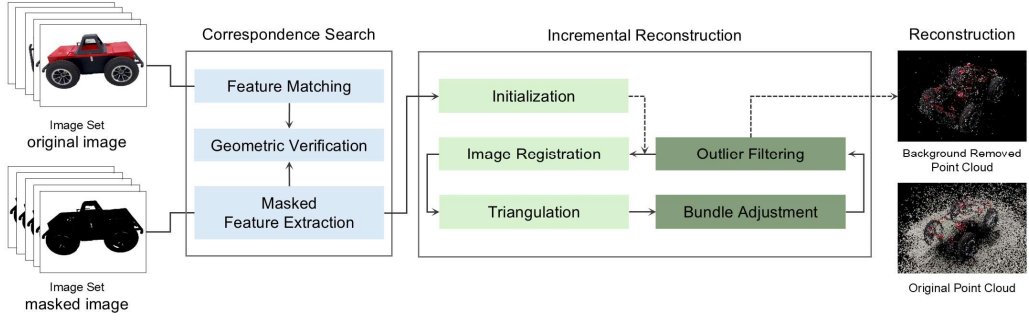

Fig. 3. Pipeline for Key Point Reduction in Structure-from-Motion. This figure demonstrates how masked feature extraction, feature matching, and geometric verification collectively ensure that only object-related key points are carried forward into incremental reconstruction, thereby producing a point cloud with minimal background content.

Before the 3D Gaussian Splatting training process, the image set undergoes structure-from-motion (SfM) to generate a 3D point cloud. Although segmentation masks are used to render the background white in the 2D images, the white background is still considered during feature extraction, leading to the formation of splats in these regions during 3D reconstruction. To address this, it is essential to remove background points from the point cloud so that splats are generated only on the object regions.

As illustrated in Figure 3, two sets of images, the original images and masked images, are utilized to mitigate the inclusion of background features. In the Correspondence Search stage, Masked Feature Extraction is applied to the binary masked images, restricting feature detection to valid object regions. This effectively reduces the number of unwanted key points arising from white background pixels. Feature Matching and Geometric Verification are then performed, retaining only object-centric correspondences across multiple views.

During Incremental Reconstruction, which encompasses initialization, image registration, triangulation, outlier filtering and bundle adjustment, the filtered key points are used to build a sparse 3D point cloud that is predominantly free of background data. Consequently, the final reconstruction focuses on the target object, preventing unwanted splats from forming in non-relevant regions when the 3D Gaussian Splatting process is applied.

### D. Attention-enhanced 3DGS Reconstruction

Even with the background points removed, standard 3DGS can struggle to preserve small or intricate details due to smoothing effects. To address this, an attention mechanism is incorporated to emphasize high-frequency regions during training and point selection as depicted in Figure 4.

First, each input image $I_{RBG}(x,y)$ is converted to grayscale by averaging the color channels. This single-channel intensity image is then convolved with two Sobel kernels, one for both the x and y directions. To compute gradients:

$$G_x(x,y) = (S_x * I)(x,y) \qquad (9)$$
$$G_y(x,y) = (S_y * I)(x,y) \qquad (10)$$

$$S_x = \begin{bmatrix} -1 & 0 & 1 \\ -2 & 0 & 2 \\ -1 & 0 & 1 \end{bmatrix} \qquad (11)$$

$$S_y = \begin{bmatrix} -1 & -2 & -1 \\ 0 & 0 & 0 \\ 1 & 2 & 1 \end{bmatrix} \qquad (12)$$

The overall gradient is computed as:

$$G(x,y) = \sqrt{G_x(x,y)^2 + G_y(x,y)^2} \qquad (13)$$

To form the attention mask $A(x,y)$, $G(x,y)$ is normalized to the range [0,1]:

$$A(x,y) = \frac{G(x,y) - \min(G)}{\max(G) - \min(G)} \qquad (14)$$

Pixels with stronger gradients meaning edges or fine textures receive higher $A(x,y)$ values, indicating areas that require greater emphasis.

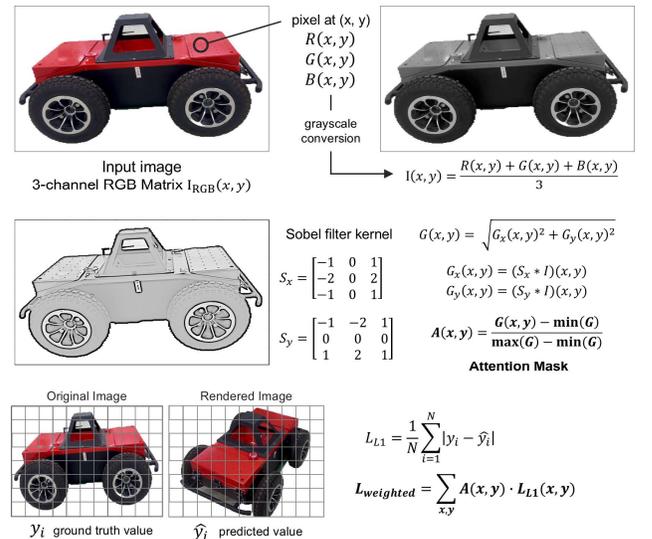

Fig. 4. This figure demonstrates how the input RGB image is converted to grayscale, processed by Sobel filters, and normalized to produce an attention mask. The mask then guides both the training loss and point pruning, preserving fine details such as edges and small structures in the 3D model.

The attention mask is applied in two key ways:

- During 3DGS training, a pixel-wise L1 loss $\mathcal{L}_{L1}$ is scaled by $A(x, y)$ to prioritize errors in high-detail regions:

$$\mathcal{L}_{weighted} = \sum_{x,y} A(x,y)|y_i - \hat{y}_i| \qquad (15)$$

This encourages the model to focus on preserving edges and fine structures.

- When projecting the point cloud onto the 2D image plane, the attention values at corresponding pixel locations guide the removal of low-confidence points. As a result, splats are more densely allocated in regions of high detail, thereby retaining critical features in the final 3D reconstruction.

## IV. RESULTS AND EVALUATION

In our experiments, the attention mechanism consistently improved the construction quality. To simulate a smart factory environment, we evaluated our approach on three distinct models of varying sizes and levels of detail: the rover model, featuring larger and more prominent elements; the hexapod model, with smaller components such as screws; and the LattePanda model, which exhibits highly intricate features. As shown in Figure 5, the rover model in View 2 demonstrates that our attention-enhanced method successfully captured a missing screw that was absent in the original 3DGS output. Similarly, for the hexapod model in View 1, the attention mechanism provided a better definition of text, enhancing the readability and structural details. The LattePanda model in View 1, which exhibits intricate details, showed improved capture of port features with our method.

Even with the background removal, the reconstructions maintained sufficient detail. In the rover model in View 1, both the original and attention-enhanced methods without background successfully captured the missing screw, and, uniquely, the attention-enhanced version also revealed holed on the top of the rover which was missed when the background was present. From View 2 of the hexapod model, the fact that the attention mechanism—when applied to background-removed images—captures fine details such as thin wires implies that the proposed approach robustly preserves small-scale features.

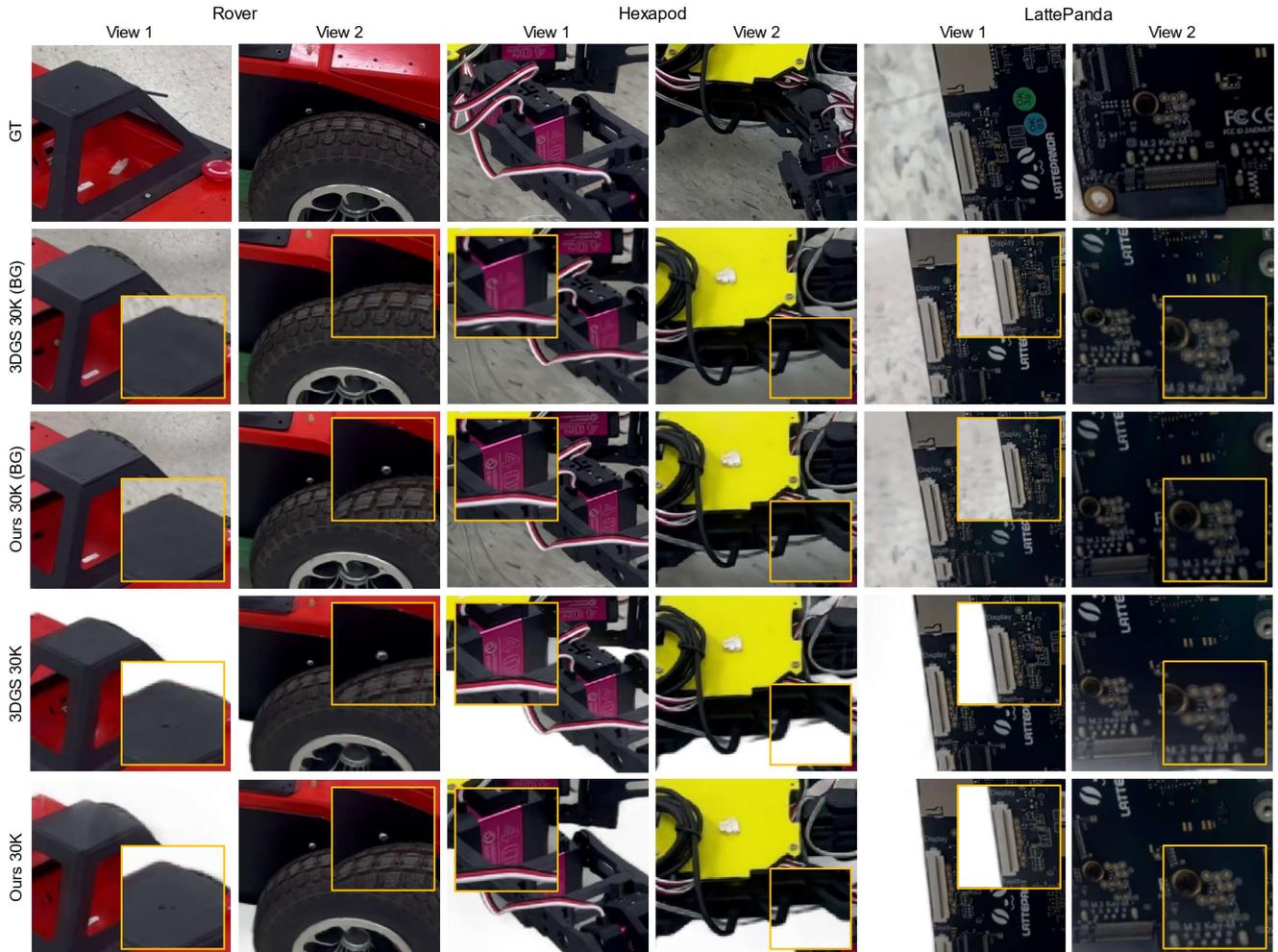

Fig. 5. Qualitative Comparison of 3D Gaussian Splatting Reconstructions. This figure presents side-by-side comparison for the rover, hexapod, and LattePanda models under two viewpoints per model. The comparisons include five sets of outputs: The Ground Truth (original input image), original 3DGS (30k iterations) with background, our method 3DGS (30k iterations) with background, original 3DGS (30k iterations) without background, and our method 3DGS (30k iterations) without background.

This indicates that even after the removal of extraneous background information, the attention-enhanced model can effectively focus on and enhance critical details, thereby overcoming potential losses in keypoint matching during the SfM process. Overall, these results imply that the integration of attention enhancement with background removal robustly improves the fidelity of 3D reconstructions, preserving critical fine details across various model complexities.

TABLE I

QUANTITATIVE RESULTS OF ROVER MODEL

| | | SSIM↑ | PSNR↑ | L1↓ | Loss↓ | FPS↑ | Time↓ | Size(MB)↓ |
|---|---|---|---|---|---|---|---|---|
| Original | 3DGS (30k) | 0.927 | 25.616 | 0.020 | 0.031 | 28.187 | 18m 23s | 338.976 |
| | Ours (30k) | 0.956 | **27.374** | 0.003 | 0.073 | 47.052 | 13m 10s | 181.477 |
| No BG | 3DGS (30k) | **0.979** | 26.368 | 0.078 | 0.096 | 42.135 | 13m 50s | 54.560 |
| | Ours (30k) | 0.976 | 27.135 | **0.003** | **0.030** | **56.980** | **11m 21s** | **29.767** |

TABLE II

QUANTITATIVE RESULTS OF HEXAPOD MODEL

| | | SSIM↑ | PSNR↑ | L1↓ | Loss↓ | FPS↑ | Time↓ | Size(MB)↓ |
|---|---|---|---|---|---|---|---|---|
| Original | 3DGS (30k) | 0.912 | 26.590 | 0.024 | 0.032 | 24.999 | 20m 11s | 332.556 |
| | Ours (30k) | 0.955 | 24.737 | **0.002** | **0.011** | 35.354 | 14m 31s | 204.258 |
| No BG | 3DGS (30k) | **0.983** | 29.163 | 0.238 | 0.250 | 46.322 | 15m 23s | 79.646 |
| | Ours (30k) | 0.974 | **29.966** | 0.007 | 0.062 | **56.088** | **10m 58s** | **39.494** |

TABLE III

QUANTITATIVE RESULTS OF LATTEPANDA MODEL

| | | SSIM↑ | PSNR↑ | L1↓ | Loss↓ | FPS↑ | Time↓ | Size(MB)↓ |
|---|---|---|---|---|---|---|---|---|
| Original | 3DGS (30k) | 0.956 | 28.878 | 0.015 | 0.021 | 33.734 | 15m 44s | 104.693 |
| | Ours (30k) | 0.969 | 29.509 | **0.001** | **0.007** | 41.522 | 11m 6s | 44.733 |
| No BG | 3DGS (30k) | **0.995** | 29.322 | 0.189 | 0.178 | 46.494 | 13m 43s | 19.840 |
| | Ours (30k) | 0.994 | **29.982** | 0.006 | 0.038 | **60.092** | **10m 45s** | **9.984** |

Across all three models – rover, hexapod, and LattePanda – the quantitative metrics (SSIM, PSNR, and L1 loss) consistently indicate the proposed approach outperforms the baseline 3DGS at 30k iterations. Background removal further improves reconstruction quality by eliminating detailed features, yielding higher SSIM and PSNR values while reducing L1 loss. These improvements highlight the synergy between attention-based detail preservation and the removal of non-essential background data.

In terms of computational performance, the frames per second (FPS) and overall training time remain comparable between the baseline and enhanced models, indicating that the additional complexity introduced by the attention mechanism does not substantially impede efficiency. The slight increase in model size is generally offset by the gains in reconstruction quality, making the trade-off practical for applications that demand high in detail. Overall, the results demonstrate that both background removal and attention enhancement contribute significantly to producing more accurate and visually coherent 3D reconstructions, regardless of object complexity or scale.

V. CONCLUSION

The proposed masked attention-based 3D Gaussian Splatting framework successfully addresses key limitations in robotic perception and object detection by combining background removal and a Sobel filter-based attention mechanism. By utilizing $U^2$-Net to isolate objects and focusing the 3DGS training on fine details, the method achieves notable improvements in visual fidelity and precision, particularly evident in the capture of small features such as screws, wires, and intricate textures. Quantitative evaluations using L1 loss, SSIM, and PSNR underscore the robustness of this approach in diverse industrial scenarios, while qualitative comparisons confirm its ability to preserve fine-scale structures. Overall, the enhanced reconstructions pave the way for more accurate object recognition and manipulation in complex smart factory environments, demonstrating the value of integrating targeted attention and background filtering into 3D vision pipelines.